\documentclass[12pt,preprint]{aastex}

\usepackage{graphics}

\def\stacksymbols #1#2#3#4{\def\theguybelow{#2}
        \def\verticalposition{\lower#3pt}
        \def\spacingwithinsymbol{\baselineskip0pt\lineskip#4pt}
        \mathrel{\mathpalette\intermediary#1}}
\def\intermediary #1#2{\verticalposition\vbox{\spacingwithinsymbol
        \everycr={}\tabskip0pt
        \halign{$\mathsurround0pt#1\hfil##\hfil$\crcr#2\crcr
                \theguybelow\crcr}}}

\begin{document}

\title{Variations of the Mid-IR Emission Spectrum in Reflection 
Nebulae\altaffilmark{1}}

\author{Jesse Bregman}
\affil{Astrophysics Branch, NASA/Ames Research Center, MS 245-6,
Moffett Field, CA 94035}

\author{Pasquale Temi}
\affil{Astrophysics Branch, NASA/Ames Research Center, MS 245-6,
Moffett Field, CA 94035; \\ \&  SETI Institute, Mountain View, CA 94043.}

\altaffiltext{1}{Based on observations with ISO, an ESA project
with instruments funded by ESA Member States (especially the PI countries:
France, Germany, the Netherlands and United Kingdom) and with the
partecipation of ISAS and NASA}

\begin{abstract}
Using spatial-spectral data cubes of reflection nebulae obtained by 
ISOCAM, we have observed a shift in the central wavelength of the 
7.7 $\mu$m band within several reflection nebulae.  This band, composed of components at 7.85 $\mu$m and 7.65 $\mu$m, shows a centroid shift from 7.75 $\mu$m near the 
edge of the nebulae to 7.65 $\mu$m towards the center of 
the nebulae as the shorter wavelength component becomes relatively stronger.    
The behavior of the 7.7 $\mu$m band center can be explained 
either by assuming that anions are the origin of the 7.85 $\mu$m 
band and cations the 7.65 $\mu$m band, or that the band center 
wavelength depends on the chemical nature of the PAHs.
The ratio of the 11.3/7.7 $\mu$m bands also changes 
with distance from the central star, first rising from the center 
towards the edge of the nebula, then falling at the largest 
distances from the star,  consistent with the 11.3/7.7 $\mu$m 
band ratio being controlled by the PAH ionization state.
\end{abstract}


\section{Introduction}

The mid-infrared emission bands observed in many environments, 
including HII regions, post-AGB (asymptotic giant branch) stars, planetary nebulae, the 
diffuse ISM of galaxies, and reflection nebulae, are still 
commonly called the unidentified infrared (UIR) bands even 
though their association with polycyclic aromatic molecules 
(PAHs) was made 20 years ago  \citep{leger,  
allamandola85} and this identification 
has since been supported by observations and theoretical calculations.  
However, identification of the emission bands with a single carrier 
has not been possible as the emission bands are likely due to a 
mixture of PAHs rather than a single molecule.  Yet, a wealth of 
observationally based information about the UIR bands has been 
collected which, for example, has allowed authors to use them to 
measure redshifts of galaxies and to differentiate between AGN 
and starburst galactic nuclei \citep{roche,genzel}.  

Trends observed 
in the detailed UIR emission spectra indicate that the molecules 
responsible for these bands vary depending on their environment 
and history.  For example, using laboratory spectra, 
\citet{allamandola99} demonstrated that the 
difference between the emission spectrum of a post-AGB star and the 
Orion Bar can be explained by varying the PAH mixture in these 
sources.  While a mixture of less stable PAH cations and neutrals 
could fit the post-AGB spectrum, more stable PAH cations with no neutrals 
were needed to fit the Orion Bar spectrum.  Since the material 
around the post-AGB star is young and the material in the Orion Bar has 
been heavily processed, they concluded that processing in the ISM 
modifies the PAH mixture.  Similar conclusions were reached by 
\citet{hony} and \citet{peeters} 
based on spectral variations they observed in ISO spectra of a 
large number of diverse sources.

The spectra of PAH neutrals and ions are quite different, with ions having stronger C-C mode (6--10 $\mu$m) emission than C-H mode (10--14 $\mu$m) emission, and neutrals having weaker C-C than C-H mode emission.  \citet{uchida} and \citet{chan} used this property to determine whether the PAH ionization state changed in reflection nebulae and the diffuse ISM.  While the reflection nebulae did show a variation of 40\% in the intensity ratio over a wide range of incident UV flux, \citet{uchida} expected a much larger variation if the PAHs changed entirely from ions to neutrals as the UV intensity decreased.


The PAH ionization state is determined by the ratio of the UV field ($G_{0}$)
to the electron density (cf.   
Bakes et al. 2001), with PAHs existing primarily as anions  for very 
low values of $G_{0}/n_{e}$ and as cations for very high values 
of $G_{0}/n_{e}$.  At intermediate values, 
PAHs will exist as a mixture of anions, neutrals, and cations.  
The ratio of the 
recombination and ionization rates determines where the transitions from cations to neutrals and neutrals to anions occur as 
$G_{0}/n_{e}$ decreases.  \citet{verstraete} have measured the ionization rate for coronene and pyrene, 
but the recombination rate (cation to neutral) has been measured only for the smallest 
PAHs \citep{abouelaziz}, and the experimental value is smaller than 
theoretical values by one to two orders of magnitude. 
Only estimates exist for the physical constants that determine the 
anion to neutral transition \citep{lepage}.  The spectra of anions and cations are very similar from 5--14 $\mu$m, so that the mid-IR spectrum in regions of very high and very low UV intensities should also be similar.

A shift of the 7.7 $\mu$m band centroid has been observed previously.  
\citet{bregman}, and \citet{cohen} noted 
that the central wavelength of the 7.7 $\mu$m band depended on 
the type of object.  Objects with freshly created PAHs 
(planetary nebulae) have a band centered near 7.85 $\mu$m, 
while objects with older material (HII regions, reflection 
nebulae) have a band centered near 7.65 $\mu$m.  They noted 
that the 7.7 $\mu$m band often contained contributions from 
both of these components.  \citet{peeters}, 
using higher spectral resolution data from the ISO database, 
confirmed these results, and could clearly distinguish the 
two components that make up the 7.7 $\mu$m band.  They also 
suggested that the PAHs showing a 7.85 $\mu$m band were pure, 
newly formed material while PAHs showing a 7.65 $\mu$m band had 
been processed in the ISM.  \citet{hony} suggested 
that as PAHs are exposed to the interstellar UV field, they are 
eroded.  Thus, it appears that as PAHs are exposed to UV 
radiation, the 7.7 $\mu$m band shifts from 7.85 to 7.65 $\mu$m.

The PAH lifecycle must be similar 
to other materials formed around stars and then ejected into the ISM 
(e.g. silicates).  Once ejected into the ISM,  PAHs are exposed to the diffuse UV field for about 10$^{7}$ 
years and chemically 
modified.  They then become part of the material 
that forms a dark cloud, where all but the most volatile components of 
the gas, including the PAHs, freeze out to form the ices observed in 
absorption in the direction of embedded protostellar sources 
\citep{sellgren, brooke, bregman3}.  
Chemical reactions should occur in the ice 
\citep{bernstein99, bernstein02}, 
again modifying the chemistry of the PAHs.  As stars burn their 
way out of the dark clouds, the PAHs become visible as 
emission from reflection nebulae.  

In this paper, using ISOCAM spectral images of reflection nebulae, 
we show that the wavelength of the 7.7 $\mu$m emission band varies 
with position in the nebula in a manner consistent with either 
variations in the ionization of the PAHs and/or with UV 
photo-processing of the PAHs.  In regions of the nebulae close to 
the exciting star, the 7.7 $\mu$m band has a weak component at 
7.85 $\mu$m which becomes stronger relative to the 7.65 $\mu$m 
component with increasing distance from the star.  
The 11.3/7.7 $\mu$m band ratio also varies with increasing 
distance from the star, first increasing and then decreasing 
in strength.  In section 2 we describe the data reduction of 
the ISOCAM images to form spatial-spectral data cubes followed 
by an analysis of the data in section 3 and a discussion in section 4.

\section{Data reduction}
A large sample of reflection nebulae was observed with the
ISOCAM camera \citep{cesarsky} on board the ISO satellite
\citep{kessler}. For the analysis presented here we selected
three reflection nebulae observed in the spectral-imaging mode with
the Circular Variable Filter (CVF). The data set is represented by
a cube in which the two spatial coordinates are defined by the 32x32
Si:Ga infrared array, and the third dimension is the spectral axis
obtained by scanning the CVF filter. The CVF is divided in two
sections that cover a spectral range from 5.14 to 9.44 $\mu$m (CVF1)
and from 9.33 to 15.1 $\mu$m (CVF2), with a spectral resolution of
$\lambda/\Delta\lambda \simeq 40$.

Two nebulae, vdB 17 and vdB 133, have been recorded by performing a full
scan of the two CVFs in both the increasing and decreasing wavelength
directions with a scale of $6\arcsec\times 6\arcsec $ per pixel. Each wavelength
was observed 11 times for vdB 133 and 8 times for vdB 17 in each scan leg,
with an elementary integration time per measurement of 2.1 seconds.
The CVF step increment was set to 1 and 2 steps respectively for vdB 133 and
vdB 17, providing an oversampling factor of 2--4 and 1--2 in the spectral
axes. At the end, the combination of the different parameter set for
each observation resulted in a total on-source integration time of
3378 and 1340 seconds, in each scan direction, for vdB 133 and vdB 17
respectively.

NGC 1333 SVS3 has been recorded with a different set of observing
parameters; the pixel scale was set to 1.5\arcsec, providing a field of view
of $48\arcsec\times 48\arcsec$. Combining the 17 exposures taken at each CVF
increment with an oversampling factor of 2--4, and a 2.1 seconds
integration time per measurements, the total observing time spent
on-source was 2696 seconds. The spectral
scan through the CVF was performed on two non-contiguous sections
of the CVF in a single scan direction (long to short wavelengths).

The data were reduced starting from the raw level, the CISP product,
using the ISOCAM Interactive Analysis processing package CIA,
(v5.0) \citep{ott}. To correct the raw data for
the dark current we used the dark model \citep{biviano}
that takes into account the observing time parameters and the
variation of the dark current both within a single revolution of
the satellite and among all the revolutions. Then, the
data were deglitched by applying a multiresolution median filter
and corrected for the transient response of the
detector. We applied the transient correction using the Fouks-Schubert
method.
Flat fielding was done using dedicated CVF zodiacal
measurements that take into account the scattered light pattern
that is produced inside the camera from reflections between
detector and CVF filters (Sauvage, private
communication).

\section{Analysis}
\subsection{Wavelength of the ``7.7'' $\mu$m band}
\begin{figure}[t]
\plotone{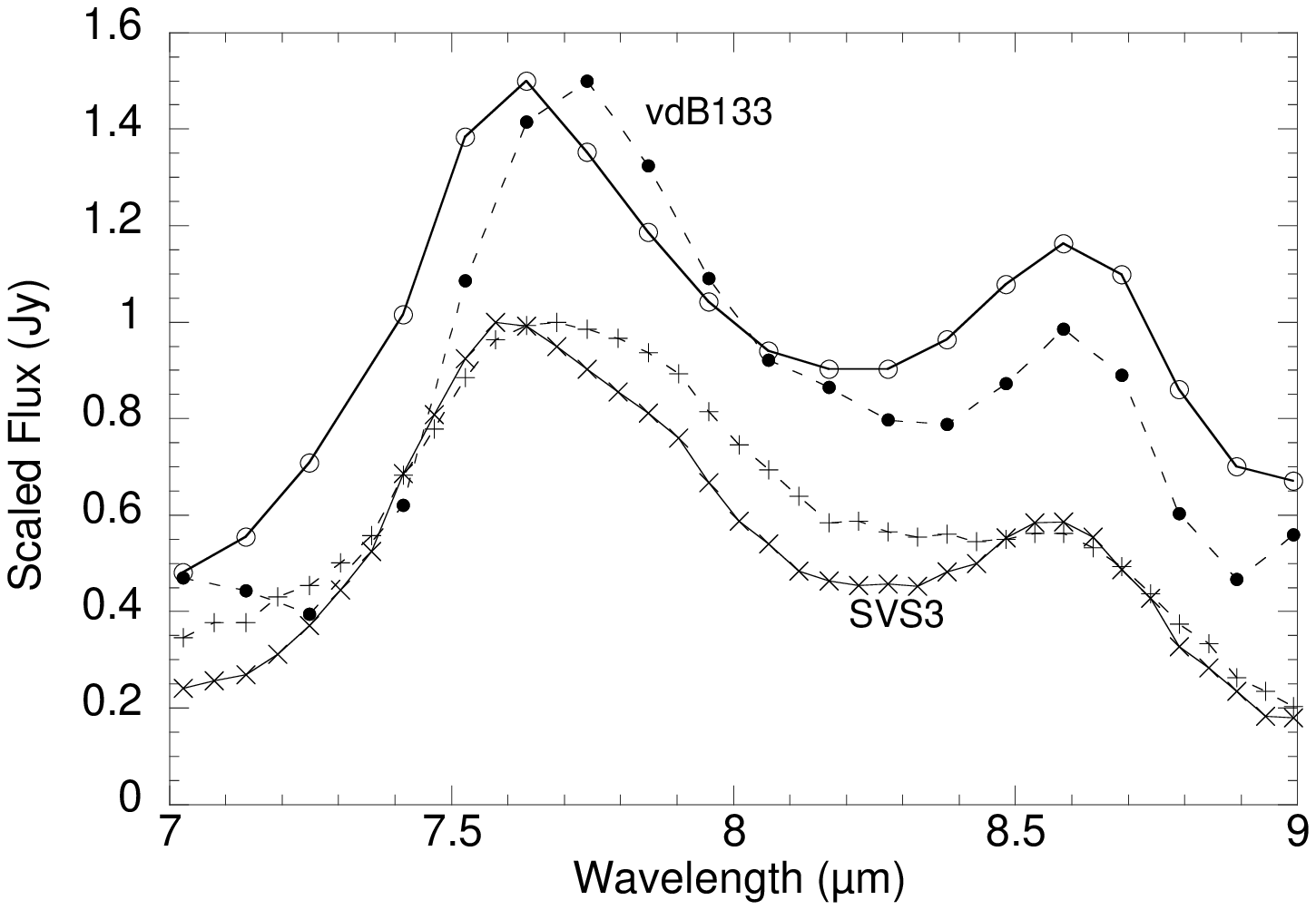}
\caption{Spectra from two different locations in each
of the two reflection nebulae vdB 133 (upper curves) and
NGC 1333 SVS3 showing the variation in central wavelength
of the 7.7 $\mu$m band with position.  The pixels with the
longer wavelength feature (and stronger 7.85 $\mu$m component)
occur farther away from the central star than the ones with
the shorter wavelength feature.  The spectra have been normalized
to one at the peaks of their emission, and a constant value of
0.5 has been added to the vdB 133 spectra for display purposes.
\label{twopix}}
\end{figure}

	Figure \ref{twopix} shows spectra from two different 
locations in the reflection nebulae vdB 133 and NGC 1333 SVS3.  
There is a clear shift in the centroid of the 7.7 $\mu$m 
emission band between the two locations in each object.  
Also in both cases, the spectrum with the shorter wavelength 
emission band is closer to the exciting star than the one 
with the longer wavelength band.  The higher resolution of 
the NGC 1333 SVS3 spectrum shows that the centroid appears 
to shift because the 7.7 $\mu$m emission band is a combination 
of two overlapping components, one peaking at approximately 
7.65 $\mu$m and a second at 7.85 $\mu$m, the same components 
observed by \citet{peeters} and 
\citet{bregman} in planetary nebulae 
(7.85 $\mu$m) and HII regions (7.65 $\mu$m).  
The 7.85 $\mu$m component is stronger towards the edges of 
the nebulae, resulting in an apparent shift of the 
7.7 $\mu$m centroid to longer wavelengths.

To explore how the center wavelength of the 7.7 $\mu$m band 
varies within these reflection nebulae, we fitted the spectrum at 
each point in vdB 133, vdB 17, and NGC 1333 SVS3 with a combination of a 
Gaussian and quadratic extending from 7.0 to 8.3 $\mu$m 
using the GAUSSFIT routine in IDL.  The quadratic fits the 
slope in the data points shortward and longward of the 
emission band to form a pseudocontinuum that the Gaussian 
rides on.  We use the center wavelength of the Gaussian 
as the centroid of the emission feature.

\subsection{Measuring feature strengths}
\begin{figure}[t]
\plotone{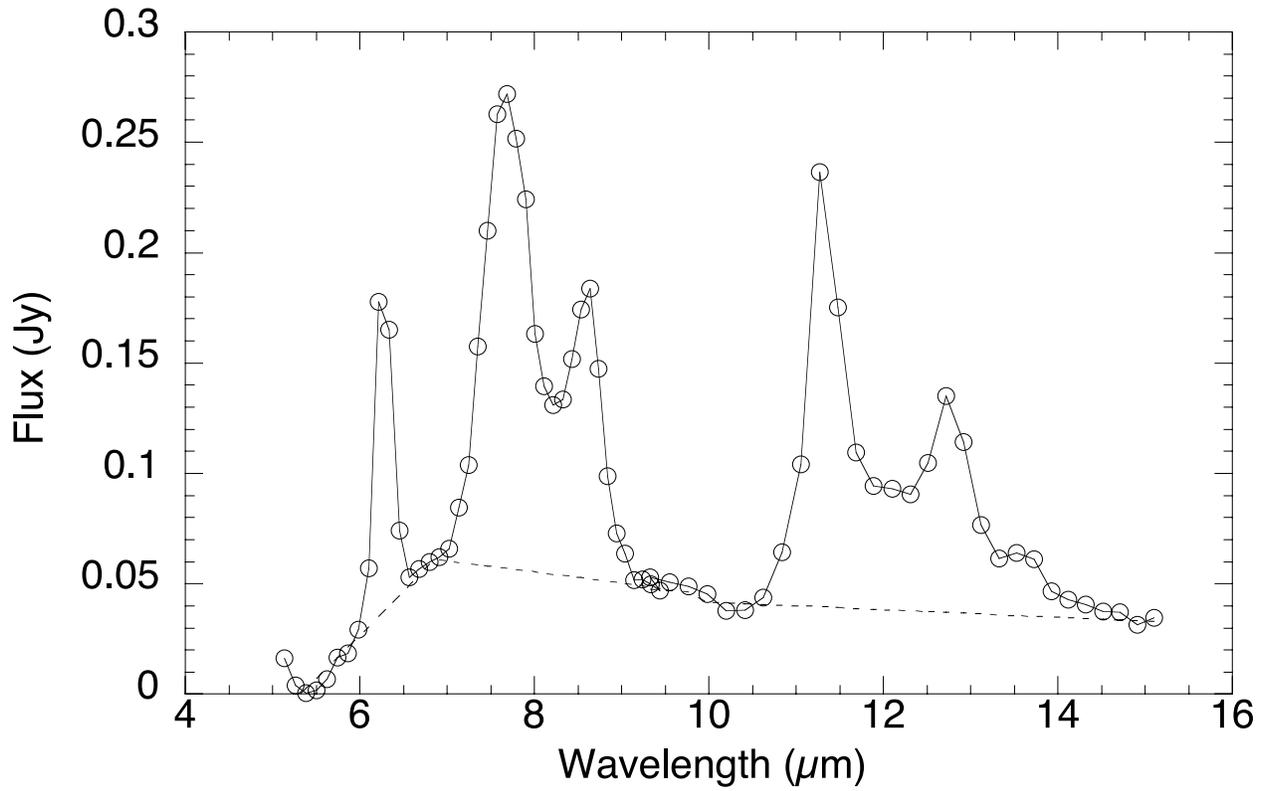}  
\caption{The spectrum of a pixel near the center of the reflection nebula vdB 17 (unfilled circles and connecting solid line) is shown with a continuum defined by straight line fits (dashed line).  The strength of the PAH emission bands are measured relative to the dashed line. \label{vdB 17spec}}
\end{figure}

The emission spectra of the reflection nebulae shown in Figure~\ref{vdB 17spec} consists of three components, narrow emission features due to PAH molecules, broad features (6--9 and 11--14 $\mu$m) due to the overlap of individual emission bands from a collection of PAHs, very large PAHs or PAH clusters, and a continuum rising to longer wavelengths attributed to very small grains \citep{sellgren1, cesarsky2, bregman2}.  Separating the components is difficult since the intrinsic band shapes are unknown.  However, it is clear from examining the data that the emission at 7, 10, and 15 $\mu$m does not decrease as quickly with increasing distance from the exciting stars as the narrow band emission.  In the fainter outer regions of the nebulae, incorrect zodiacal light subtraction also contributes to this "continuum".  

Since we are interested in the response of PAH ionization to a changing UV field, we have chosen to only include emission above a continuum level defined by fitting straight line segments between data points centered at 5.8, 7.0, 9.3, 10.0, 
and 14.9 $\mu$m.  This procedure removes both the contribution from very small grains and any residual spectral 
structure in the sky background that could affect the accuracy of 
the band intensities, primarily in the low intensity regions 
of the nebulae.  

\begin{figure}
\epsscale{.60}
\plotone{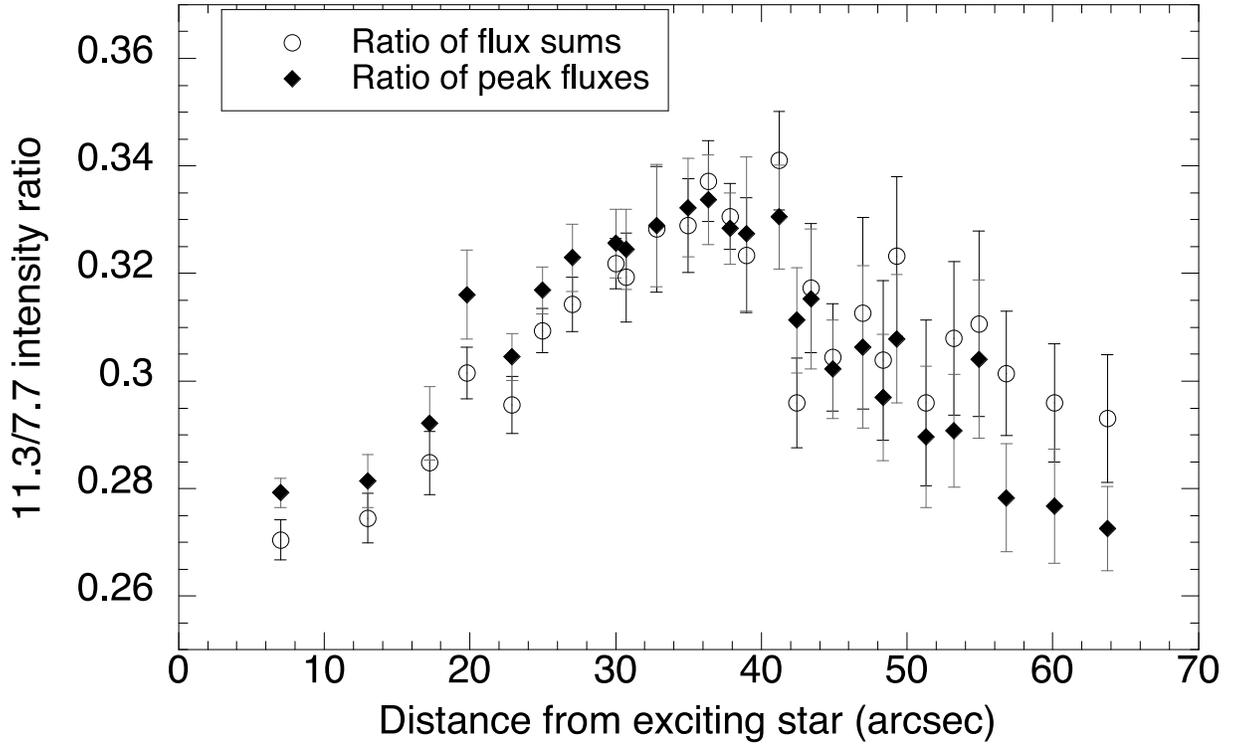}
\caption{The intensity ratio of the 11.3 to 7.7 $\mu$m emission bands is shown as a function of distance for vdB 17.  The open circles show the ratio for the integrated band strengths while the filled diamonds show the ratio for the band peak intensities.  Each point is an average of ten pixels and the error bars show the standard error for each group of pixels. \label{vdB 17bands}}
\end{figure}

In vdB 17, we compared band ratios derived by 
integrating the band intensities of the 7.7 and 11.3 $\mu$m 
bands with peak band intensity ratios.  The integrated 7.7 $\mu$m band includes all of the flux above the continuum between 6.8 and 9.1 $\mu$m, and the integrated 11.3 $\mu$m band includes the flux between 10.4 and 11.9 $\mu$m.  Figure~\ref{vdB 17bands} shows a comparison of the 11.3/7.7 $\mu$m intensity ratio as a function of distance from the exciting star.  Each point is an average of 10 pixels, and the error bars show the standard errors. Both methods produce curves with similar shapes for the 11.3/7.7 $\mu$m intensity ratio, and the mean error for both methods is the same (about 3\%).  The peak band intensities are a 
factor of 3.35 higher than the integrated band ratios since 
the 7.7 $\mu$m band is much broader than the 11.3 $\mu$m band. The integrated 7.7 $\mu$m band includes a band of unknown origin centered near 8 $\mu$m that increases in strength towards the edges of the nebulae, adding to the integrated flux \citep{uchida}.  Since both methods produce the same result,  we have chosen to use peak intensity ratios to avoid possible contamination by the 8 $\mu$m feature.

\subsection{Correlation of the wavelength of the ``7.7'' 
$\mu$m band with other parameters}

For each location in the nebulae, there is a strong correlation 
between the distance from the exciting star  and the wavelength 
of the 7.7 $\mu$m band.  This is shown in Figure \ref{7.7 vs 
distance} as a function of the incident UV field, which is 
also a function of distance from the exciting star.  The UV 
field intensity is generally denoted as $G_{0}$, 
where $G_{0}$=1 is the average value of the interstellar UV 
field and has a value of 1.6$\times$10$^{-3}$ ergs s$^{-1}$ cm$^{-2}$.  
Determining the UV field intensity at each point in a nebula is not 
necessarily straightforward since there is generally no information 
about the relative positions (in 3-D space) of the nebula and the 
exciting star.  In most cases, authors have chosen to use the 
projected distance on the sky as the true distance for calculating 
the UV field intensities.  

\begin{figure}
\epsscale{.70}
\plotone{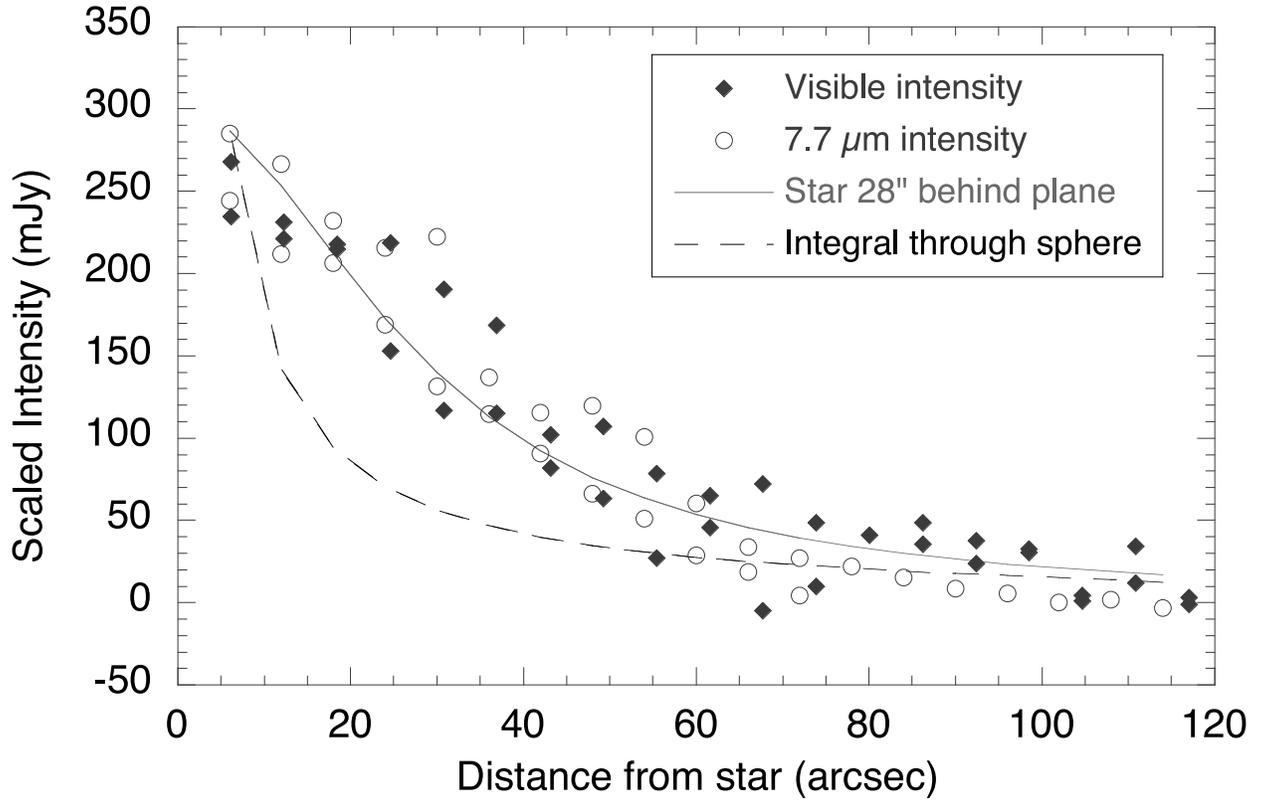}
\caption{The radial brightness profile of a north-south cut
through vdB 17 in both visible light (from the Palomar digital
sky survey) and at the peak of the 7.7 $\mu$m band.  The dashed
line is the intensity profile expected for a uniform density
sphere with the exciting star in the center, while the solid
line is the intensity profile expected for a star behind a
scattering layer. \label{radial profile}}
\end{figure}

In the case of vdB 17, where the exciting 
star appears to be centered on the nebula, 
\citet{uchida} assumed that the exciting star was embedded within 
the nebula, so that the UV intensity dropped off as r$^{-2}$ 
where r is the projected distance on the sky. 
\citet{witt} point out that the brightest reflection 
nebulae have their exciting stars either within or behind the 
nebula, so the assumption that the star is within the nebulosity 
in vdB 17 seems reasonable.  However, this assumption can be tested 
by examining the scattered light intenstiy of the nebula as a 
function of distance.  In Figure \ref{radial profile} we show 
the radial brightness profile of a north-south cut through vdB 17 
in both visible light (from the Palomar digital sky survey) and 
at the peak of the 7.7 $\mu$m band.  If the star is embedded 
within the nebulosity, and the nebula is spherical and has 
uniform density, then the light intensity should decrease with distance as shown 
by the dashed line in Fig. \ref{radial profile}, which does not fit the data.  If the nebula 
is highly flattened and the star is within the nebula, the 
falloff will be even faster.  However, if the star is behind 
the nebula by a distance corresponding to 28\arcsec on the sky 
(2$\times$10$^{17}$ cm), and the nebula is flat and of uniform density, 
then the intensity profile can be fitted quite well (solid line 
in Fig. \ref{radial profile}).  Using this geometry gives a much 
smaller range for $G_{0}$ as a function of projected distance than 
that assumed by \citet{uchida}.  Both geometries give 
low values of $G_{0}$ far from the star, but the geometry with the 
star behind the nebulosity also has relatively low values for the UV 
intensity close to the center of the nebula.
	
	For vdB 133, the density is not uniform and we do not have a 
method for determining the relative geometry of the star and nebula.  
However, the star is some distance to the side of the nebulosity 
so that the calculated distance from the star to any point in the 
nebula is not as sensitive to the assumed geometry as in the case of 
vdB 17.  For our discussion, we will take the projected distance of the 
star to the nebula as the true distance and use that for calculating 
the incident UV field.  In NGC 1333 SVS3, the star appears centered in 
the nebulosity, but the density does not appear to be uniform.  
In this case, we have chosen to use the projected distance as the 
true distance, although this could introduce an error for points 
close to the star.

\begin{figure}
\epsscale{1.0}
\plotone{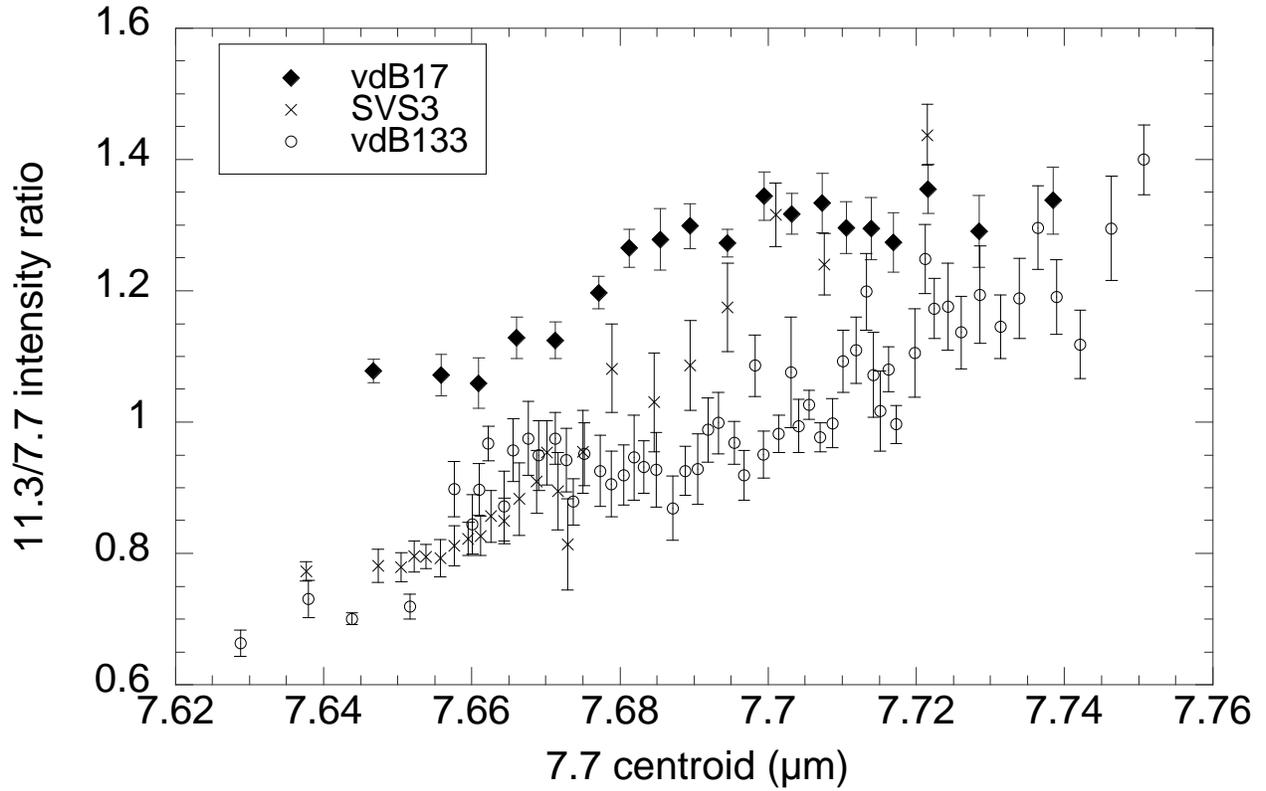}
\caption{The correlation of the intensity ratio of the 11.3
and 7.7 $\mu$m bands compared to the center wavelength of
the 7.7 $\mu$m band  for vdB 133, vdB 17, and SVS3.  Each point
shown is the average of 10 original data points, and the error
bars are the standard error in each group of 10. \label{ratio cor 77}}
\end{figure}

The wavelength of the 7.7 $\mu$m band also correlates well 
with the intensity ratio of the 11.3 and 7.7 $\mu$m bands 
(shown in Fig. \ref{ratio cor 77} for vdB 17, vdB 133, and SVS3).

\section{Discussion}
Since the UV intensity changes by two orders of magnitude within each reflection nebula, and since the ratio of the UV intensity to the electron density, $G_{0}/n_{e}$, determines the PAH ionization state, it is necessary to estimate both $G_{0}$ and $n_{e}$ within the nebulae before we can examine whether changes in the PAH ionization as a function of position within the nebulae can account for the observations.  In the following sections, we will first discuss how we determined $G_{0}$ and $n_{e}$ for the nebulae, then model the observations in two different ways.

\subsection{Expected ionization state of the PAHs}	

The PAH ionization state within a reflection nebula is a function of the ratio of the UV intensity to the electron density.  The UV intensity is a function of the nebular geometry and the spectral type of the illuminating star, and the electron density can be determined roughly for these 
reflection nebulae by estimating the density of the gas within 
the nebulae.  \citet{joblin96b} give an electron 
density of 15 e$^{-}$ cm$^{-3}$ in NGC 1333 SVS3 based on a 
C/H ratio of 3$\times$10$^{-4}$, assuming that all of the electrons 
would come from ionization of neutral carbon, and an estimate 
of the gas density based on H$_{2}$ emission \citep{joblin96a}.  
\citet{owl} derive a density 
for SVS3 of 2$\times$10$^{4}$ (n$_{e}$=6 cm$^{-3}$) based on a PDR 
model and far IR measurements centered at the far IR peak 
intensity.

In the reflection nebula vdB 17 (also in the NGC 1333 region), 
the density can be estimated from the extinction through the 
nebula.  \citet{racine} gives E(B-V)=0.6 for vdB 17, 
so that A$_{v}$=1.8 magnitudes.  Using the canonical column density value of 
N$_{H}$/A$_{v}$=1.9$\times$10$^{21}$ cm$^{-2}$ magnitude$^{-1}$ gives 
N$_{H}$=3.4$\times$10$^{21}$ cm$^{-2}$ along the line of sight to the 
star in vdB 17.  We can derive a lower limit to the density 
by assuming that the nebula is spherical.  The nebula has an 
apparent diameter of Å100\arcsec, corresponding to 
7.5$\times$10$^{17}$ cm at a distance of 500 pc 
\citep{strom}.  The average density is then 9.1$\times$10$^{3}$ cm$^{-3}$, and the electron 
density (also assuming that the electrons are from carbon 
ionization) will be 2.7 cm$^{-3}$.  We already have stated that 
the exciting star is 28\arcsec behind the scattering nebulosity, 
so it is very likely that the thickness of the nebula along the 
line of sight is substantially less than 100\arcsec, perhaps by 
an order of magnitude, and that the electron density is 
correspondingly higher.   
\citet{martini} derive a density of 
$\sim$1$\times$10$^{4}$ cm$^{-3}$ (n$_{e}$=3 cm$^{-3}$) 
for vdB 17 based on molecular hydrogen line ratios.  
\citet{warin} derive a density of 3$\times$10$^{3}$ cm$^{-3}$ 
to 5$\times$10$^{4}$ cm$^{-3}$ (n$_{e}$=0.9--15 cm$^{-3}$) based on CO line 
observations.  Thus, while the electron density in vdB 17 is 
uncertain, it is probably in the range of 1--10 cm$^{-3}$, 
and we will use 1 cm$^{-3}$ for the following discussion.

For vdB 133, \citet{li} derive an electron density 
of 0.3 cm$^{-3}$ based on a model fit to the PAH emission spectrum.  
We can estimate the average density in the same manner as we did 
for vdB 17, using the reddening to the nebula and its apparent size.  
\citet{racine} gives E(B-V)=0.65, the apparent projected extent 
of the nebulosity in the infrared is 120\arcsec, and it is at a distance 
of 1400 pc \citep{humphreys}.  Assuming that the nebula 
has the same extent along the line of sight as in the plane of the 
sky gives an average density of 1.4$\times$10$^{3}$ cm$^{-3}$ and n$_{e}$=0.4 cm$^{-3}$.  
While both of these values are uncertain, we do not have any 
better methods of estimating the density in vdB 133, and will 
adopt a value of n$_{e}$=0.3 cm$^{-3}$ for the following discussion.
	
\begin{figure}
\plotone{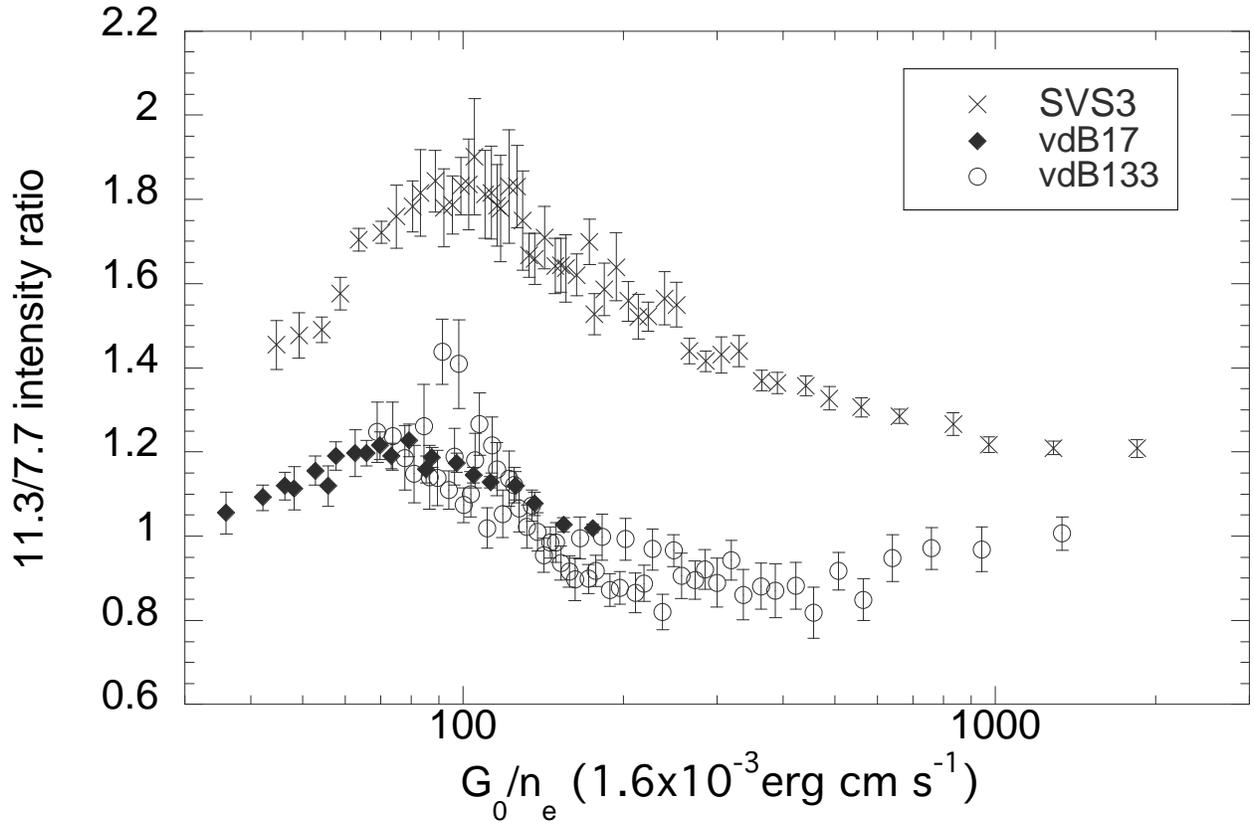}
\caption{The 11.3/7.7 $\mu$m intensity ratio plotted as a function
of $G_{0}/n_{e}$ show the effects of PAH ionization.
The SVS3 data points have been shifted upwards by 0.5 for
display purposes.  For vdB 17 n$_{e}$=1 cm$^{-3}$, for vdB 133
n$_{e}$=0.3 cm$^{-3}$, and for SVS3 n$_{e}$=15 cm$^{-3}$. \label{pah ions}}
\end{figure}

Using the value of $G_{0}$ for SVS3 given by 
\citet{joblin96a} and those given by 
\citet{uchida} for vdB 17 and vdB 133, we can calculate 
$G_{0}/n_{e}$ as a function of distance in the nebulae.  
Fig. \ref{pah ions} shows the 11.3/7.7 $\mu$m 
band ratio as a function of $G_{0}/n_{e}$ for the three 
nebulae.  Each data point is an average of ten pixels 
and the error bars show the standard error of each group 
of ten.  A constant value of 0.5 has been added to the SVS3 
data for display purposes since it overlaps the data from the 
other two nebulae.  Qualitatively, the observed trend is as 
expected for the 11.3/7.7 $\mu$m band ratio since as 
$G_{0}/n_{e}$ increases, PAHs should transition from anions 
through neutrals to cations.  \citet{bakes} 
show that the strength of the 7.7 $\mu$m band is a slowly changing function 
of $G_{0}/n_{e}$, with a value 30\% larger at $G_{0}/n_{e}$=10 
than at $G_{0}/n_{e}$=1$\times$10$^{4}$.  Their result is also 
consistent with our observation that the 7.7 $\mu$m band 
tracks the scattered light intensity in vdB 17 
(Fig. \ref{radial profile}) which requires that most of the 
observed variation in the 7.7 $\mu$m band intensity be due 
to re-emission of absorbed radiation rather than abundance 
variations.  The 11.3 $\mu$m band is strongest in neutral PAHs, 
so that the observed peak in the 11.3/7.7 $\mu$m band ratio 
at $G_{0}/n_{e}$=100 can be interpreted as the value of 
$G_{0}/n_{e}$ where there is a maximum in the abundance of 
neutral PAHs.  This is  about a factor of four lower than 
calculated by \citet{bakes}, which can simply 
be the result of Bakes et al. using too large a value for 
the PAH$^{+}$ recombination coefficient.  
\citet{chan} also preferred a low value for 
the recombination coefficient  based on their measurements 
of relative PAH band intensities  in the diffuse ISM.  Thus, 
as $G_{0}/n_{e}$ decreases from 1000 to 100, the cation to 
neutral ratio drops.  Below $G_{0}/n_{e}$=100, anions are 
forming and the 11.3/7.7 $\mu$m band ratio decreases.  If this 
interpretation is correct, then it appears that PAHs transition 
almost directly from cations to anions with only a narrow 
range of  $G_{0}/n_{e}$ where neutral PAHs are present, and that 
the abundance of neutral PAHs is never very high.  This could 
also explain why the PAH emission spectrum doesn't change very 
much from object to object as the anion and cation spectra 
are very similar in the mid-IR.  However, even if the abundance of neutral PAHs is never very high, they can contribute significantly to the 11-14 $\mu$m emission since these bands are much stronger in neutral PAHs than in PAH ions.
		
\begin{figure}
\plotone{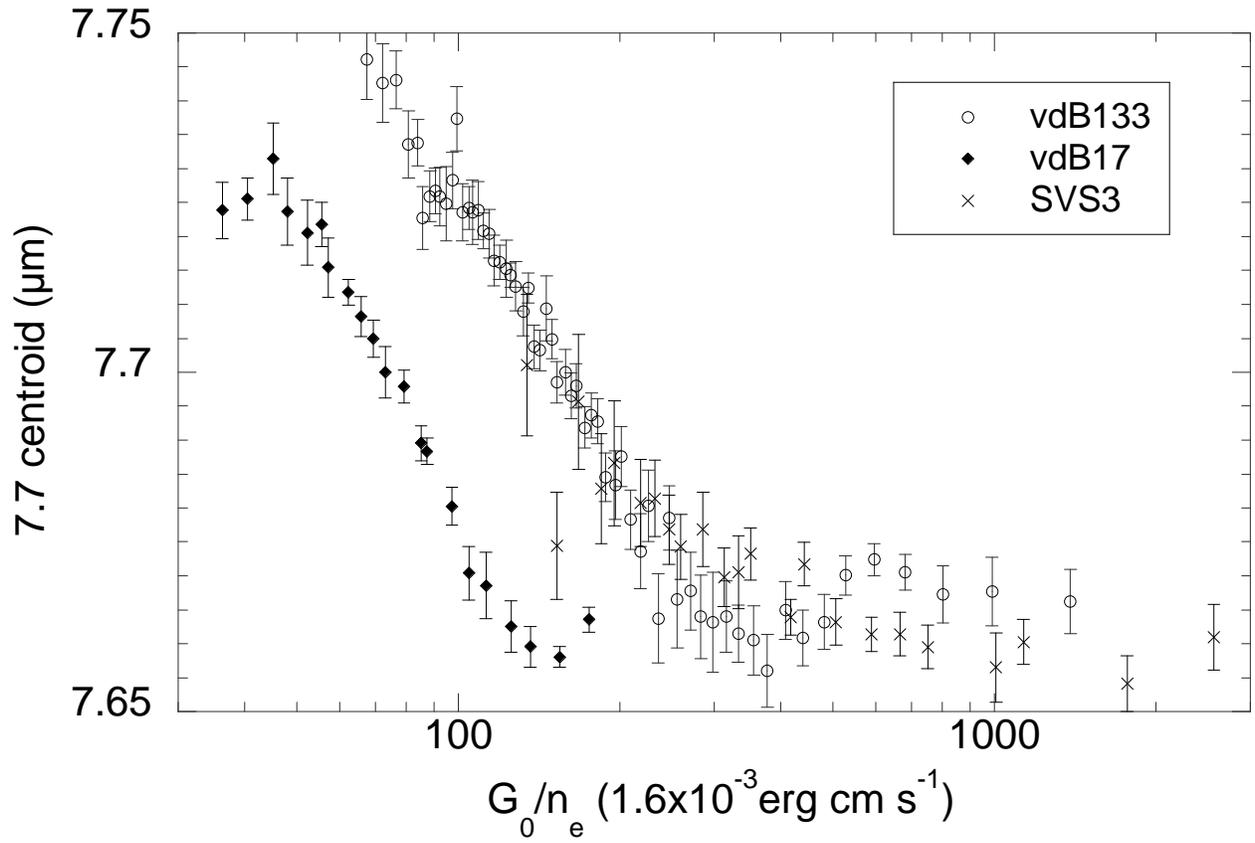}
\caption{The center wavelength of the 7.7 $\mu$m band is plotted
as a function of $G_{0}/n_{e}$.  For vdB 17 n$_{e}$=1 cm$^{-3}$, for vdB 133
n$_{e}$=0.3 cm$^{-3}$, and for SVS3 n$_{e}$=15 cm$^{-3}$. \label{7.7 center vs gone}}
 \end{figure}

Fig. \ref{7.7 center vs gone} shows the 7.7 $\mu$m centroid plotted as a function 
of $G_{0}/n_{e}$ where each point plotted is the average 
of 10 pixels.  The coincidence of the vdB 133 and SVS3 curves 
and the near coincidence (within a factor of two in 
$G_{0}/n_{e}$) of the vdB 17 curve, can be explained if the 
process that causes the band centroid to shift is a chemical 
equilibrium process.  An example of such a process is detailed 
in the next section.

\subsection{Modeling the PAH Emission Process 1: The Wavelength of 
the 7.7 $\mu$m Band as an Indicator of the Anion/Cation Ratio }

\begin{figure}
\plotone{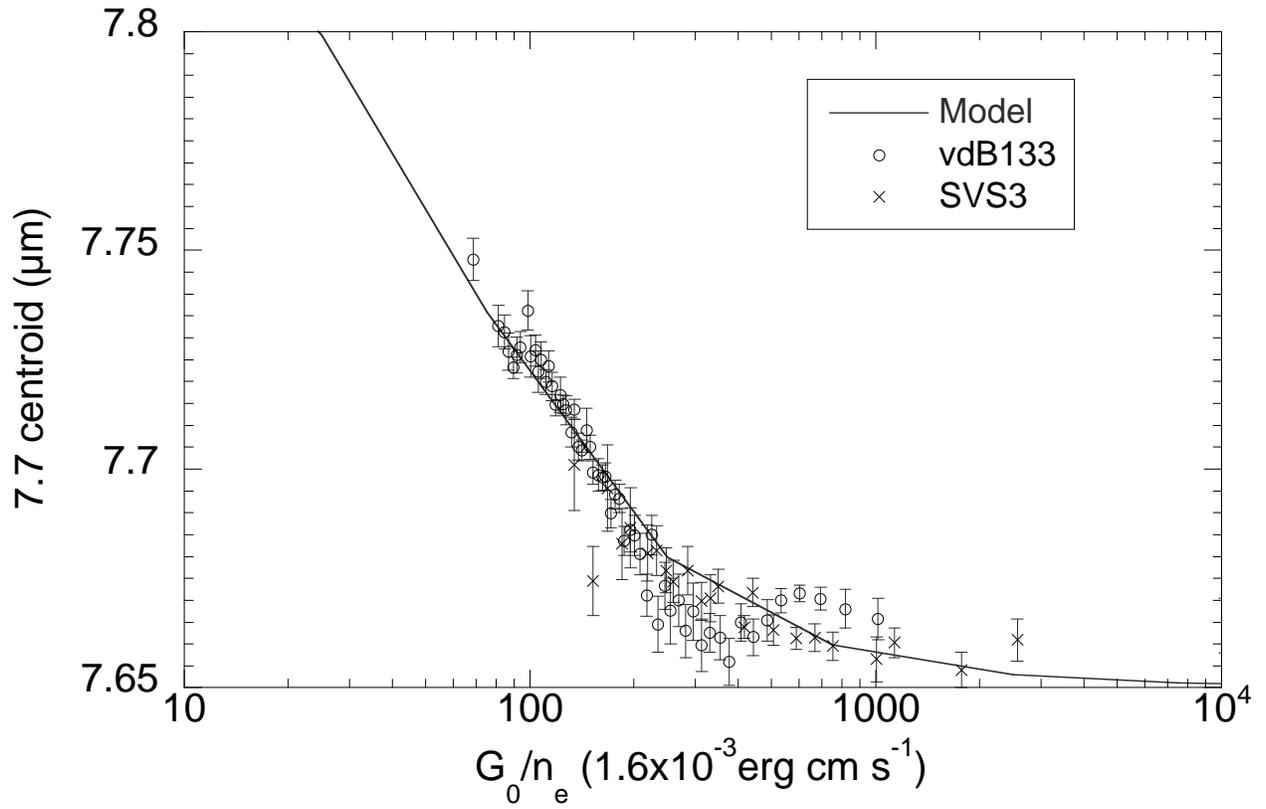}
\caption{The 7.7 $\mu$m band center as a function
of $G_{0}/n_{e}$ is fitted by a model (solid line) in which
cations emit at 7.65 $\mu$m and anions at 
7.85 $\mu$m. \label{7.7 center model}}
\end{figure}

To explain the observed behavior, we can construct a simple model in 
which we will assume that anions have an emission band at 7.85  
$\mu$m while cations have an emission band at 7.65 $\mu$m.  Then 
we calculate the ionization balance between anions and neutrals 
and neutrals and cations separately assuming ionization equilibrium.  
Following \citet{verstraete}, the ionization 
balance between neutrals and cations can be expressed as
\begin{equation}
n_{n}r_{n}G_{0}=n_{+}n_{e}k_{+}
\end{equation}
where $n_{n}$ is the number of neutral PAHs, $r_{n}$ is the ionization 
rate coefficient for neutral PAHs, $n_{+}$ is the number of singly 
ionized PAHs, $n_{e}$ is the electron density, and $k_{+}$ is the 
recombination coefficient for singly ionized PAHs.  The ionization 
balance between anions and neutrals can be described by a similar 
equation.  Rearranging the equation gives an expression for the 
fraction of PAHs that are singly ionized ($f_{+}$).
\begin{equation}
f_{+}=\frac{G_{0}}{G_{0}+n_{e}K_{+}}
\end{equation}
where $K_{+}=k_{+}/r_{n}$.  The fraction of anions ($f_{-}$) is
\begin{equation}
f_{-}=\frac{n_{e}}{n_{e}+G_{0}K_{-}}
\end{equation}
where $K_{-}=r_{-}/k_{n}$, $r_{-}$ is the rate for removing an 
electron from anionic PAHs, and $k_{n}$ is the attachment rate 
for electrons to neutral PAHs.
The centroid of the 7.7 $\mu$m band ($\lambda_{c}$) is then
\begin{equation}
\lambda_{c}=\frac{7.65f_{+}+7.85f_{-}}{f_{+}+f_{-}}
\end{equation}
For a PAH molecule with 80 carbon atoms, \citet{verstraete} 
give $r_{n}=3.2\times10^{-8}$ sec$^{-1}$.  Using this value, we then 
adjusted $k_{+}$ and the ratio $r_{-}/k_{n}$ until the 
7.7 $\mu$m centroid calculated as a function of $G_{0}/n_{e}$ 
fitted the data (solid line, Fig. \ref{7.7 center model}).  
The fit requires a value for $k_{+}$ of $3.2\times10^{-6}$ cm$^{-3}$sec$^{-1}$, a factor 
of two less than that given by  
\citet{verstraete}, and a value of $1.8\times10^{-2}$ cm$^{3}$ 
for $r_{-}/k_{n}$, a factor of three less than the estimate given by \citet{lepage}.  Using these parameters, we 
calculated $f_{+}$ and $f_{-}$ as a function of $G_{0}/n_{e}$, 
and the neutral fraction ($f_{n}$) as $1-f_{+}-f_{-}$.

\begin{figure}
\plotone{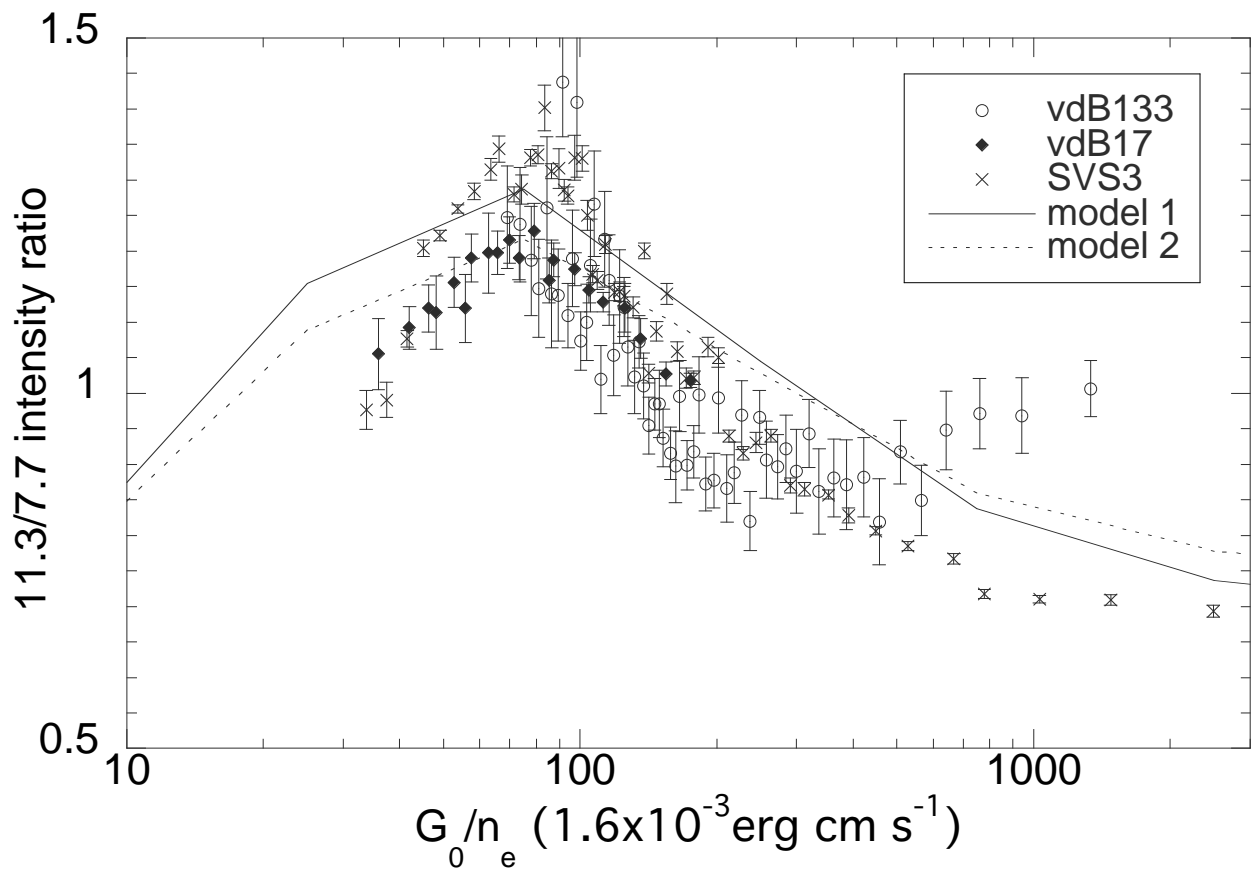}
\caption{The 11.3/7.7 $\mu$m band ratio as a function
of $G_{0}/n_{e}$ is shown with a model (solid line)
consistent with the intrinsic band ratios used by
\citet{bakes} and a model (dotted line) with
the intrinsic 11.3 $\mu$m band strength for neutral PAHs
reduced by 20\%. \label{bakes model}}
\end{figure}

Given the PAH ionization state as a function of $G_{0}/n_{e}$, 
we can also calculate the 11.3/7.7 $\mu$m ratio as a function 
of $G_{0}/n_{e}$ if we know the intrinsic 7.7 and 11.3 $\mu$m 
band strengths for anions, neutrals, and cations.  The intrinsic 
band strengths are the average for the entire PAH population 
weighted by the abundance of each PAH, and will depend on the 
PAH mix present in each object and on the spectrum of the 
exciting UV field.  For this calculation, we assume that the 
PAH mix does not change within an object, and that the intrinsic 
band strengths are the same as those calculated by  
\citet{bakes}.  Then, the 11.3/7.7 $\mu$m ratio is
\begin{equation}
\frac{I_{11.3}}{I_{7.7}}=\frac{0.59f_{-}+4.0f_{0}+1.2f_{+}}{1.2f_{-}+0.62f_{0}+1.6f_{+}}
\end{equation}
and is shown as the solid line in Fig. \ref{bakes model} along with 
the data from the reflection nebulae.  In addition, the dashed line 
shows the effects of reducing the intrinsic 11.3 $\mu$m band intrinsic 
strength for neutrals by 20\%.  The agreement between the model and 
the data is fairly good, especially since we did not adjust any of 
the intrinsic band strengths to create a better fit (for example, 
a small decrease in the intrinsic 11.3/7.7 $\mu$m band ratio for 
cations would result in a better fit for points with $G_{0}/n_{e}$ 
greater than 200 and a small increase for neutrals would give a 
better fit near $G_{0}/n_{e}=100$).  The three points plotted at 
the lowest values of $G_{0}/n_{e}$ do not fit the model very well, 
but we have made the assumption here that the density is constant 
within the entire nebula and that there is no UV extinction 
internal to the nebula.

\subsection{Modeling the PAH Emission Process 2: The Wavelength of 
the 7.7 $\mu$m Band as an Indicator of PAH Processing}	

A second model that could fit the data assumes that the position of the 7.7 $\mu$m band depends on the physical nature of the emitting PAHs rather than on the ionization state of the PAHs. \citet{hony} suggested that PAH molecules are eroded by exposure to the interstellar UV field, and these physical changes are reflected in the relative strengths of the C--H out-of-plane bending modes.  It is possible that the same physical changes cause the 7.7 $\mu$m band to shift from 7.85 $\mu$m, as observed in planetary nebulae, to 7.65 $\mu$m, as observed in HII regions.  The outer regions of reflection nebulae, where the band centroid is closer to 7.85 $\mu$m, would contain a larger fraction of less eroded PAHs than the regions closer to the exciting star where the UV intensity is higher.

\begin{figure}
\plotone{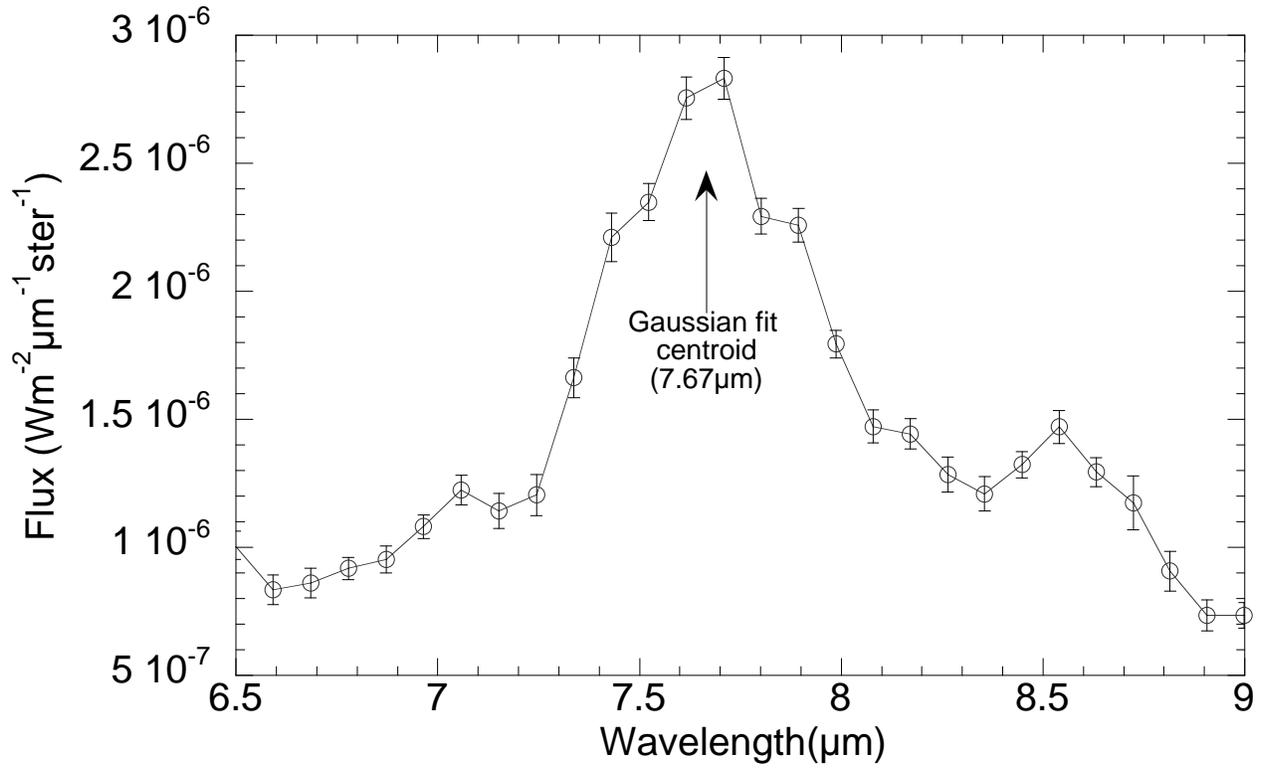}
\caption{The 7.7 $\mu$m band in the diffuse ISM (from ISOPHOT)
has a central wavelength of 7.67 $\mu$m, similar to the processed
PAHs found in HII regions. \label{diffuse ism}}
\end{figure}

The molecular clouds from which reflection nebulae are born are formed from the gas and dust of the ISM, and in the scenario we are considering here, the PAHs leaving these clouds have a 7.7 $\mu$m band centered near 7.85 $\mu$m.  We can determine the wavelength of the 7.7 $\mu$m band for PAHs entering these clouds 
by examining the spectra of PAHs in the 
diffuse ISM.  \citet{mattila} used the ISOPHOT 
spectrometer (PHT-S) to take spectra of the galactic plane along 
lines of sight that avoided bright IR emission and optically 
bright stars.  We averaged spectra taken at (l,b) of (-45,0), (-30,0), (-15,0), (-15,1), (-5,0), and (+30,0), all regions with obvious PAH emission, to produce the spectrum shown 
in Fig. \ref{diffuse ism}.  Using these data, we determined the position of the 
7.7 $\mu$m band in the diffuse ISM, finding that its centroid 
occurs at 7.67 $\mu$m, the same wavelength (within the precision 
possible with the low resolution spectra from ISOPHOT) as found 
in HII regions and reflection nebulae, implying that material 
entering HII regions and the dark clouds from which reflection 
nebulae form have 7.7 $\mu$m bands centered near 7.65 $\mu$m.

\begin{figure}
\epsscale{1.0}
\plotone{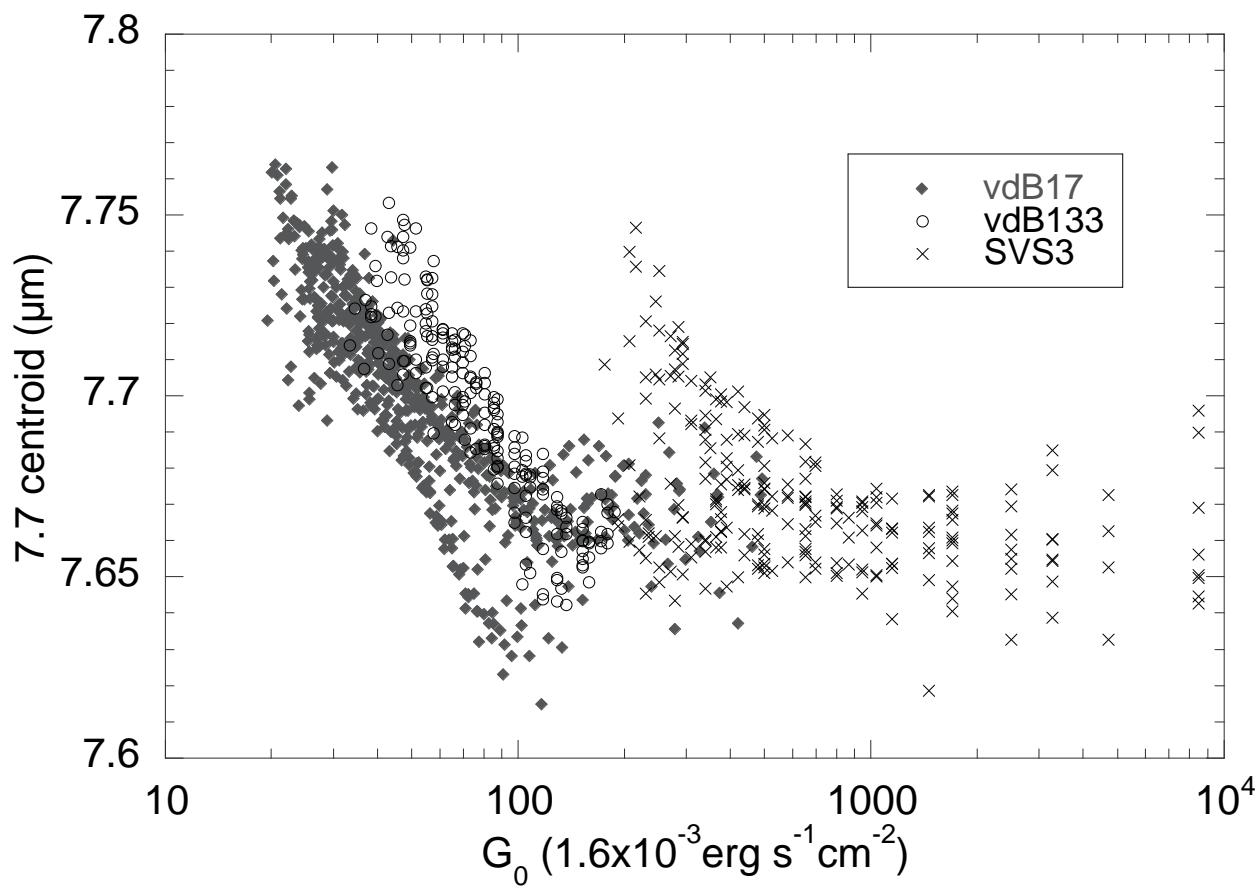}
\caption{For vdB 17, vdB 133, and NGC 1333 SVS3 there is a strong
correlation between the wavelength of the 7.7 $\mu$m band and the
incident UV field (or equivalently the distance from the exciting
star).  The center of the 7.7 $\mu$m band shifts shortward as the
UV field increases, leveling off at a wavelength of around 7.65
$\mu$m, indicating that the 7.85 $\mu$m component no longer is
contributing to the feature for larger UV exposures.
For SVS3, $G_{0}$ has been divided by 10 (shifted to the left) for plotting 
purposes. \label{7.7 vs distance}}
\end{figure}

In this model, PAHs are chemically altered during the 
time that the PAH molecules reside within the dark cloud, and molecules are formed that are similar 
to the fresh PAHs observed around planetary nebulae.  It is these 
chemically altered molecules that have a strong 7.85 $\mu$m 
band and are most abundant at positions in the reflection nebulae that 
are farthest from the exciting stars, and thus have suffered the least 
amount of cumulative UV exposure.  Nearer to the exciting stars, the 
PAHs are processed as they are during residence in the ISM, and 
the 7.65 $\mu$m band becomes more prominent.  
Fig. \ref{7.7 vs distance} 
shows that as the UV field intensity increases, the band centroid 
shifts to shorter wavelengths.  For vdB 133, the band centroid shift 
occurs as $G_{0}$ increases from 20 to about 100, but then remains 
constant at about 7.65 $\mu$m for larger values of $G_{0}$, while 
in SVS3, the transition to a constant value occurs at about 
a $G_{0}$ of 700.  This is the behavior expected if the PAHs are fully 
reprocessed by the UV exposure for intensities above $G_{0}$=100 
(for vdB 133, or 700 for SVS3) since once all the material is converted 
to the form with a strong 7.65 $\mu$m band, it will not appear to shift 
any further.  The behavior in vdB 17 mirrors that in vdB 133.
If the above explanation is correct, then we expect that the UV exposure 
that PAHs experience before they are ejected into the ISM plus the 
exposure within the ISM should be comparable to the exposure they 
suffer in these reflection nebulae.  Assuming a residence time in the 
ISM of 1$\times$10$^{7}$ years, and where by definition $G_{0}$=1, the total 
UV exposure can be expressed as the product of these two values, 
or 1$\times$10$^{7}$ in units of $G_{0}$ years.  The UV exposure of a 
PAH before it is ejected from a planetary nebula is more difficult 
to estimate.  PAHs are likely formed during the mass-loss phase of 
carbon-rich stars and then modified by exposure to UV radiation to 
produce the PAH spectrum characteristic of planetary nebulae.  
The 7.85 $\mu$m component is usually stronger than the 7.65 $\mu$m 
component in planetary nebulae, but there is a wide range observed 
in the ratio of the two components.  For example, in 
BD +30$^{\circ}$3639 the 7.85 $\mu$m component is dominant while in 
NGC 7027, the two components are almost of equal strength 
\citep{peeters}.  The emitting PAHs are at the edges 
of the ionized region (within the surrounding photo-dissociation region), 
and if they have experienced the same UV field for the lifetime of the 
nebulae, then the total UV exposure in the two objects is 
similar ($\sim$5$\times$10$^{7} G_{0}$ years).  However, the actual exposure 
is certainly less than this value since in the past the ionized region 
was smaller and the PAHs that are now visible were shielded by material 
closer to the central star.  Thus, the PAHs that are present in the ISM 
likely have a cumulative UV exposure of a few $\times$10$^{7}$ $G_{0}$ years.  
	
To determine the UV exposure of PAHs in the reflection nebulae, it is 
necessary to estimate the age of the nebulae.  Both vdB 17 and SVS3 
are in the NGC 1333 complex, which includes a star cluster with an 
age of 1--2$\times$10$^{6}$ years \citep{lada}.  
\citet{joblin96a} adopted an age of 1$\times$10$^{5}$ 
years for SVS3.  The reflection nebulae must be somewhat 
younger than the stars since they formed only after the stars 
broke out of the dark cloud.  Also, there is no reason why both 
nebulae should have the same age.  SVS3 is more compact than 
vdB 17 and probably has a higher density, perhaps indicating that 
it is younger.  Using 1$\times$10$^{5}$ years for the age of SVS3 gives 
a range for the cumulative UV exposure of the PAHs of 1$\times$10$^{7}$ 
to 1$\times$10$^{9}$ for $G_{0}$=100 to 10,000, with the PAHs showing no 
change in the 7.7 $\mu$m centroid position above $G_{0}$=700, 
corresponding to a UV exposure of 7$\times$10$^{7}$ $G_{0}$ years.  
For vdB 17, an age of 1$\times$10$^{5}$ gives an exposure of 
from 2$\times$10$^{6}$ to 2$\times$10$^{7}$ $G_{0}$ years for $G_{0}$=20 to 200, and an 
exposure ten times higher if the age is 1$\times$10$^{6}$ years.  
In both of these objects, the cumulative UV exposure of the 
PAHs is comparable to the UV exposure of PAHs in the diffuse ISM.  
The near coincidence of vdB 17 and vdB 133 in Fig. \ref{7.7 vs distance} 
then implies that these two objects have similar ages.  
	
\begin{figure}
\plotone{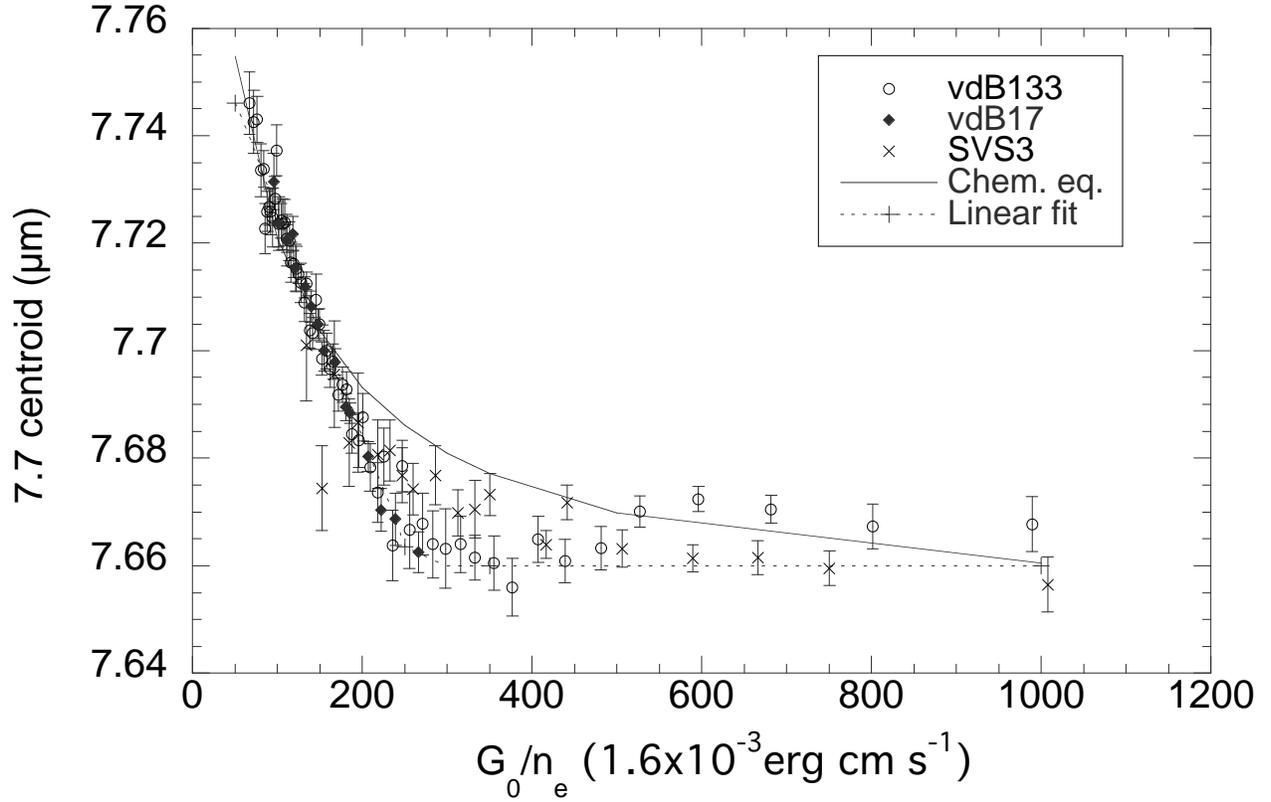}
\caption{The 7.7 $\mu$m band centroid as a function
of $G_{0}/n_{e}$ can be fitted either with a straight line
(dotted) or one characteristic of a chemical equilibrium
reaction (solid line).  The data for vdB 17 have been shifted
by about a factor of two to match the vdB 133 
data. \label{straight line fit}}
\end{figure}

Fig. \ref{straight line fit} shows two model fits to 
the 7.7 $\mu$m centroid as a function of $G_{0}/n_{e}$.  If UV processing of the PAHs causes the 
wavelength shift, then the 
shift could be linear with exposure ($G_{0}\times$time) if it is due 
to an irreversible reaction (dotted line) such as loss of side groups \citep{joblin96a}.  
However, in this case the co-alignment of the curves when 
plotted vs $G_{0}/n_{e}$ would not be expected unless density is linearly 
related to the age of the nebulae.  If the shift is due to a reversible 
reaction (ie. loss of H), then it should scale as $G_{0}/n_{e}$ and 
follow the solid curve.  Thus, while this curve may not fit as well as 
a linear fit, the co-alignment of the data from the three nebulae is a 
consequence of the equilibrium reaction process.  In this plot, the 
vdB 17 data has been shifted to larger $G_{0}/n_{e}$ by about a factor 
of two, which is equivalent to lowering the density by a factor of two.

\subsection{Comparison of the models}
Changes in the ionization balance of PAHs provides a good explanation for the changing ratio of the 11.3 to 7.7 $\mu$m features, and this explanation is supported by models \citep{bakes} and from fitting laboratory spectra to astronomical sources \citep{allamandola99}.  Thus, the shift of the 7.7 $\mu$m feature could be simply explained if anionic PAHs have a feature at 7.85 $\mu$m while cationic PAHs have a feature at 7.65 $\mu$m.  However, there is no evidence, either from theoretical spectral calculations or from laboratory spectra, that the PAH ionization state has an effect on the position of the 7.7 $\mu$m band.  This may simply mean that spectra of the appropriate PAH ions do not yet exist.

\begin{figure}
\epsscale{.70}
\plotone{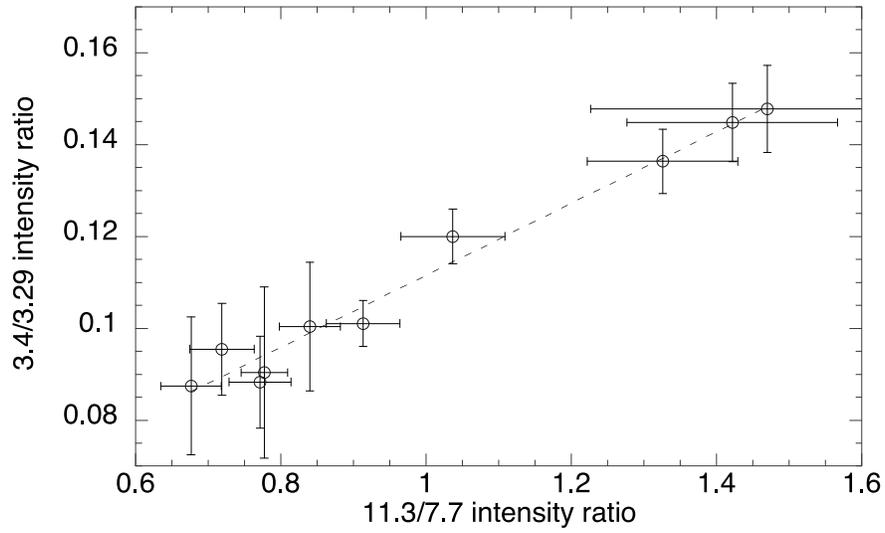}
\caption{In the region south of the exciting star in NGC 1333 SVS3, the 3.4/3.3  $\mu$m band intensity ratio correlates with the 11.3/7.7 $\mu$m intensity ratio.  The 3.4/3.3 $\mu$m intensity ratio is an indicator of the aliphatic to aromatic C-H bond ratio, while the 11.3/7.7 $\mu$m intensity ratio is an indicator of PAH ionization.  \label{aliphatic_aromatic}}
\end{figure}

The alternate explanation, that the shift of the 7.7 $\mu$m band is due to chemical modification of the PAH population, is supported in two ways.  First, \citet{hony} show that as PAHs are exposed to UV radiation, the solo, duo, and trio out-of-plane C-H modes in the 11--13 $\mu$m region change in a manner that can be explained by erosion of the edges of the molecules. Secondly, \citet{joblin96a} and \citet{sloan97} showed that in SVS3, the 3.4/3.3 $\mu$m band intensity ratio increases with increasing distance from the exciting star.  We show the data from \citet{sloan97} as a function of the 11.3/7.7 $\mu$m band intensity ratio (an indicator of PAH ionization) in Fig. \ref{aliphatic_aromatic}.  \citet{joblin96a} showed that the 3.4 $\mu$m band is due to -CH$_3$ sidegroups on PAH molecules that are lost upon exposure to UV radiation with a corresponding decrease in the 3.4/3.3 $\mu$m band intensity ratio.  Thus, Fig. \ref{aliphatic_aromatic} shows that chemical evolution of PAH molecules and the ionization state of the molecules can be well correlated.  We conclude that the present data does not rule out either explanation.

\section{Conclusions}
We have observed a shift in the central wavelength of the 
7.7 $\mu$m band within three reflection nebulae.  Previous 
observations had established that the 7.7 $\mu$m band has a 
strong component near 7.85 $\mu$m in fresh PAH material (e.g. around 
planetary nebulae), while PAHs that have been exposed to the 
UV of the ISM for long periods (e.g. HII regions, diffuse ISM) are 
dominated by a band near 7.65 $\mu$m.  In the reflection nebulae 
studied here, the wavelength of the feature occurs at longer 
wavelengths at points distant from the exciting stars in the 
nebulae, and at progressively shorter wavelengths for points 
closer to the stars.  We also observe that the 11.3/7.7 $\mu$m 
band ratio changes as a function of $G_{0}/n_{e}$ in a manner 
consistent with changes in PAH ionization.  

The shift of the 7.7 $\mu$m band position can be explained either 
as a result of changes in the anion to cation ratio (assuming that 
anions have a strong band at 7.85 $\mu$m and cations have a 
strong band at 7.65 $\mu$m) or as an indicator of UV processing of 
the PAH mixture.  In the second scenario, PAHs enter a dark cloud 
with a 7.65 $\mu$m band center and are chemically processed on grain 
surfaces, causing a shift of the band to 7.85 $\mu$m.  Upon exposure 
to UV radiation, the PAHs are modified back to the form they had in the 
diffuse ISM and the band shifts back to 7.65 $\mu$m.

\begin{acknowledgements}
      We wish to thank Lou Allamandola, Max Bernstein, 
Doug Hudgins, and Greg Sloan for their helpful comments.
\end{acknowledgements}

\clearpage

\end{document}